\documentclass[11pt,twoside]{article}
\usepackage{atmp}
\usepackage{latexsym,amsmath,amssymb,graphicx}
\copyrightnotice{2002}{6}{619}{642}

\def\inbar{\,\vrule height1.5ex width.4pt depth0pt}

\newcommand{\CO}{{\cal O}}
\newcommand{\IR}{\relax{\rm I\kern-.18em R}}
\newcommand{\IT}{\relax{\rm I\kern-.18em T}}
\newcommand{\hx}{{\widehat X}}
\newcommand{\IP}{{\relax{\rm I\kern-.18em P}}}
\newcommand{\CY}{{\cal Y}}
\newcommand{\CF}{\cal F}
\def\IZ{\mathbb {Z}}
\def\IF{{\relax{\rm I\kern-.18em F}}}
\def\mo{\overline{M}}
\def\om{\overline{M}}
\def\tr{{\rm tr}}
\def\bz{{\overline Z}}
\def\by{{\overline Y}}
\def\bcy{{\overline{\cal Y}}}
\def\inj{\hookrightarrow}
\newcommand{\of}{{\overline f}}

\def\IC{{\relax\hbox{$\inbar\kern-.3em{\rm C}$}}}
\begin{document}

\title{Geometric Transitions \\ and Open String Instantons} 
\author{D.-E. Diaconescu,$^1$ B. Florea,$^2$ and A. Grassi$^3$}
\address{$^1$ Department of Physics and Astronomy,
Rutgers University,\\
Piscataway, NJ 08855-0849, USA}
\addressemail{email: duiliu@physics.rutgers.edu}
\vspace*{.1in}
\address{$^2$Mathematical Institute, University of Oxford,
\\24-29 St. Giles', Oxford OX1 3LB, England}
\addressemail{email: florea@maths.ox.ac.uk}
\vspace*{.1in}
\address{$^3$ Department of Mathematics, University of
Pennsylvania,\\Philadelphia, PA 19104-6395, USA}
\addressemail{email: grassi@math.upenn.edu}
\url{hep-th/0205234}  

\pagestyle{myheadings}
\markboth{GEOMETRIC TRANSITIONS AND OPEN STRING INSTANTONS}{DIACONESCU, FLOREA AND GRASSI}
\noindent
\noindent
We investigate the physical and mathematical structure of a new class
of geometric transitions proposed by Aganagic and Vafa. The distinctive
aspect of these transitions is the presence of open string instanton
corrections to Chern-Simons theory.
We find a precise match between open
and closed string topological amplitudes applying a beautiful idea proposed
by Witten some time ago. The closed string amplitudes are reproduced from
an open string perspective as a result of a fascinating interplay of
enumerative techniques and Chern-Simons computations.  
\newpage

\section{Introduction}
\label{intro}

Open topological strings on Calabi-Yau manifolds have been the subject of
much recent activity in the context of geometric transitions
and enumerative geometry \cite{AAHV}-\cite{B},\cite{GViii}-\cite{HIV},\cite{IK}-\cite{KL},\cite{LMi}-\cite{LM},\cite{LS}-\cite{OV},\cite{RS}-\cite{sts}.
In particular, Aganagic and Vafa \cite{AViv} found a new class of large $N$
geometric dualities which yield very interesting predictions for open
topological string amplitudes on noncompact Calabi-Yau threefolds.
Recall that the original geometric transition discovered by Gopakumar
and Vafa \cite{GViii} relates open strings on a deformed conifold
to closed strings on the blow-up of the same conifold singularity.
As shown by Witten many years ago \cite{EWii}, the topological open string
on a deformed conifold is equivalent to $U(N)$ Chern-Simons theory on
the vanishing $S^3$ cycle.
The new transitions of \cite{AViv} predict a similar relation for a more
complex geometric set-up in which the open string theory is corrected
by instanton effects.
Such situations have been anticipated by Witten
in \cite{EWii}, where the instanton corrections have been elegantly interpreted
as non-local Wilson loop operators in the Chern-Simons action. The Wilson
loops in question are boundaries of holomorphic discs (or higher
genus bordered Riemann surfaces) interpreted as knots in $S^3$.
Then the computation of the open string free energy reduces to a fascinating
combination of open string enumerative geometry and perturbative
Chern-Simons theory.

In this paper we show that these techniques can be successfully
applied to the large $N$ duality proposed in \cite{AViv}, resulting in
a precise match between open and closed string amplitudes. The
computation of open string amplitudes in this background entails
two different aspects. One has first to compute the instanton
corrections to the Chern-Simons effective action, which is
essentially a problem in open string enumerative geometry
\cite{GZ,KL,LS}. In fact, a detailed (and quite involved)
analysis shows that in this particular model all these corrections
are generated by multicovers of a single rigid disc. The
corresponding instanton expansion has been predicted by Ooguri and
Vafa in \cite{OV} and computed explicitly by Katz and Liu \cite{KL} and Li
and Song \cite{LS} . The next step is a Chern-Simons computation, in
which the instanton corrections are treated as non-local Wilson
loop perturbations \cite{EWii}. We show that the final result is in
perfect agreement with the closed string free energy, provided
that one takes into account a certain correction to the duality
map. A remarkable aspect of this correspondence is that a priori
the closed and open string instanton expansions exhibit different
multicover contributions. This discrepancy is miraculously
accounted for by the perturbative Chern-Simons corrections, so
that in the end we obtain a precise match.

The paper is structured as follows. Section two is a brief review
of the geometric construction of \cite{AViv}, and the duality predictions.
In section three we review and develop the computation of closed string
amplitudes of \cite{AViv}, finding a complete expression for the closed
string free energy. Section four is devoted to open string amplitudes
and the duality map, and we conclude with a series of technical considerations
on open string morphisms in section five.

\section{The Transitions}
\label{thetrans}

The starting point of our discussion is the first model considered in
\cite{AViv}, namely a toric noncompact Calabi-Yau threefold $X$ defined by
the toric quotient $\left(\IC^5\setminus F\right)/(\IC^*)^2$

\begin{equation}\label{eq:toricA}
\begin{array}{rrrrr}
X_{0} & X_{1} & X_{2} & X_{3} & X_{4}\cr
1 & 0 & -1 & 1 & -1\cr
-2 & 1 & 0 & 0 & 1\cr
\end{array}
\end{equation}
\\where the disallowed locus is $F = \{X_0=X_3=0\}\cup \{X_1=X_3=0\}\cup
\{X_1=X_4=0\}$.This toric quotient can be equivalently described as a
symplectic quotient $\IC^5//U(1)^2$ with the moment maps

\begin{equation}\label{eq:momentA}
\begin{aligned}[b]
|X_0|^2 -|X_2|^2 + |X_3|^2 -|X_4|^2& = &\hbox{Re}(s)\cr 
-2|X_0|^2 + |X_1|^2+|X_4|^2& =& \hbox{Re}(t)\cr
\end{aligned}
\end{equation}
where $(s,t)$ are complexified K\"ahler parameters with
$\hbox{Re}(s)>0$, $\hbox{Re}(t)>0$.
Note that for $t=s=0$, the quotient (\ref{eq:momentA})
is a singular variety described in terms of invariant
polynomials
$x= -X_2X_3,\ y=X_0X_1X_4,\ u = X_0X_3X_4^2,\ v=X_0X_1^2X_2$
by\\
\begin{equation}\label{eq:signA}
uv+ xy^2=0.
\end{equation}
This singularity admits several distinct crepant resolutions corresponding
to nonzero values of $(\hbox{Re}(s),\hbox{Re}(t))$, which parameterize the
extended K\"ahler cone.
As usual, each such resolution corresponds to a different triangulation
of the toric diagram, and different resolutions are related by flops.
The triangulation corresponding to (\ref{eq:momentA}) is represented in fig. 1.

\begin{figure}[htp]
     \centering
     \scalebox{1}{\includegraphics{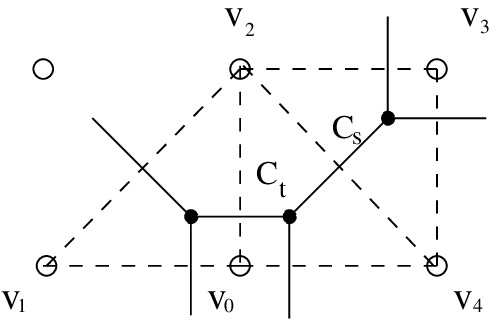}}
     \caption{Triangulation of the toric variety (\ref{eq:momentA}) with
$\hbox{Re}(s)>0$, $\hbox{Re}(t)>0$}\label{hha}
    
\end{figure}

Using this description, it easy to see that the exceptional locus consists
of two smooth rational curves $C_s, C_t$ with normal bundles
\begin{equation}\label{eq:normbundles}
N_{C_s/X} = \CO(-1)\oplus \CO(-1), \qquad
N_{C_t/X} = \CO\oplus \CO(-2)
\end{equation}
As suggested by the notation, the two curves have volumes
$\int_{C_s}J = \hbox{Re}(s)$,
$\int_{C_t}J=\hbox{Re}(t)$. Moreover,
$C_s$ is rigid while $C_t$ moves in a one parameter family on $X$.
A very useful and intuitive description of $X$ can be achieved by
representing it as a (topological) $T^2\times \IR$ fibration over $\IR^3$
\cite{MG,LV}. The discriminant of this fibration is the
planar graph represented with continuous lines in fig. \ref{hha}.
In this framework,
the curves $C_s, C_t$ can be described as topological $S^1$ fibrations
over certain line segments ending on the edges of $\Gamma$.
We will not give more details here since this is standard material
\cite{MG,LV}.
The extremal transition considered in \cite{AViv} consists of a contraction of
$C_s$ on $X$, followed by a smoothing of the resulting nodal singularity.
Let us denote by $\hx$ the singular variety obtained in the process.
As discussed in \cite{AViv}, $\hx$ can be realized as a partial resolution
of the singularity (\ref{eq:signA}) obtained by blowing-up the plane $u=y=0$

\begin{equation}\label{eq:blowupA}
u\rho=y\lambda,~~uv+xy^2=0.
\end{equation}
There are two coordinate patches $U_1$, $U_2$
on the total space of the blow-up $(x,y,v,\lambda)$ and
$(x,u,v,\rho)$, with transition functions

\begin{equation}\label{eq:transfct}
x=x,~~v=v,~~u=y\lambda,~~\rho=\frac{1}{\lambda}.
\end{equation}
This shows that the blow-up $Z$ is isomorphic
to the total space of the rank three bundle $\CO\oplus \CO\oplus \CO(-1)$
over $\IP^1$.
The local equations of the proper transform $\hx$ in the two
patches are
\begin{equation}\label{eq:proptransfA}
v\lambda + xy=0,~~v+xu\rho^2=0.
\end{equation}
Note that there is a conifold singularity left in
the first patch, which can be smoothed out by deforming the
equations as follows
\begin{equation}\label{eq:defA}
v\lambda + xy =\mu,~~v+xu\rho^2 =\mu\rho
\end{equation}
where $\mu\in \Delta^*$ is a complex
deformation parameter on the unit disc.
Throughout this paper we
will work at some fixed arbitrary value of $\mu$, which can be
assumed real and positive without loss of generality\footnote{If
$\mu=|\mu|e^{i\phi}$ with $\phi\neq 0$, we can reduce to
$\mu'=|\mu|\in\IR_{+}$ by a change of coordinates $x'=xe^{i\phi}$,
$v'=ve^{i\phi}$.}. Let $\CY/\Delta$ denote the family
parameterized by $\mu$.
The generic fiber
$Y_\mu$ is a smooth noncompact Calabi-Yau threefold, and the central fiber
is isomorphic to $\hx$. Moreover, standard surgery arguments \cite{HC} show that
the nonzero Betti numbers of a generic smooth fiber are $b_2(Y_\mu) = 1$,
$b_3(Y_\mu)=1$.
The third homology is generated by a vanishing 3-sphere
$L_\mu\subset Y_\mu$, which is the fixed point set of the local antiholomorphic
involution $\iota :U_1\rightarrow U_1$, $\iota :(x,y,v,\lambda)\rightarrow ({\overline y},
{\overline x}, {\overline \lambda}, {\overline v})$. 
Note that $L_\mu$ is lagrangian with respect to any symplectic form
$\omega$ on $Z$ such that $\omega|_{U_1}$ is odd under the involution $\iota$.
Such a symplectic form can be constructed as follows.
We can think of
$Z$ as a direct product $Z=W\times \IC^2$, where
$W$ is the total space of $\CO(-1)$ over $\IP^1$. We have coordinates
$(x,v)$ on the $\IC^2$ factor, and local coordinates $(y,\lambda)$
and $(u,\rho)$ on $W$. The local antiholomorphic involution $\iota$
is of the form $\iota=(\kappa,\kappa^{-1})$, where $\kappa$ is a
map $\kappa : \IC^2 \rightarrow U_1\cap W$, $\kappa:(x,v)\rightarrow
({\overline y},{\overline \lambda})$.
Pick a symplectic K\"ahler form
$\eta$ on $W$, and let $\eta_1$ be the restriction of $\eta$
to $U_1\cap W$. Then $\eta'=\kappa^*\eta_1$ defines a
symplectic form on $\IC^2$,
and we can take $\omega = \eta -\eta'$ to be the desired
symplectic form\footnote{To be more precise, let $\pi_{1,2}$ denote
the projections from $Z=W\times \IC^2$ onto the two factors.
Then $\omega = \pi_1^* \eta -\pi_2^*\eta'$.} on $Z$. From 
now on we fix such a symplectic form on $Z$.
The second homology group $H_2(Y_\mu,L_\mu; \mathbb{Z})=\mathbb{Z}$ is generated by
a holomorphic disc $D$ in $Y_\mu$ with boundary on $L_\mu$, which
can be constructed as follows. Let $D$ be the disc $|t|\geq
\mu^{1/2}$ in a projective plane $\IP^1$ with homogeneous
coordinates $[t_1, t_2]$, and affine coordinates $t=t_1/t_2$,
$t'=t_2/t_1$. In local coordinate patches, the embedding map
$f:D\rightarrow Y_\mu$ is given by
\begin{equation}\label{eq:holdiscA}
\begin{aligned}[r]
&U_1:&  &\lambda(t)=t,& & v(t)=\frac{\mu}{t},& & x(t)=0,& & y(t)=0\\
&U_2:&  &\rho(t')=t',& & v(t')=\mu t',& & x(t')=0,& & u(t')=0.
\end{aligned}
\end{equation}

It is easy to check that this local
description of the map is compatible with the transition functions
(\ref{eq:transfct}) and that the boundary of $D$ is mapped to $L_\mu$. Note
that $D$ is preserved by the antiholomorphic involution.
 It will be proven
in section five that $D$ is an integral generator and that there
are no holomorphic curves on $Y_\mu$. Therefore we can define an
open string K\"ahler parameter as the symplectic area of the disc

\begin{equation}\label{eq:symarea}
t_{op} = \int_D f^*\omega.
\end{equation} 

A deformation argument
to be detailed in section five shows that $t_{op}=t$ at classical
level.

In the context of geometric transitions, we have to consider an
open string topological theory defined by wrapping $N$ D-branes
on the sphere $L_\mu$. This theory is well defined since
$L_\mu$ is lagrangian. Then, large $N$ geometric duality \cite{AViv}
predicts a relation between closed string free energy on $X$ and
open string free energy on $Y_\mu$ of the form
\begin{equation}\label{eq:dualityAi}
{\CF}^{cl}(\widehat{t},\widehat{s},g_s)|_{\widehat{s}=i\lambda}={\CF}^{op}(\widehat{t}_{op},\lambda,g_s)
\end{equation}
where $\lambda=Ng_s$ is the 't Hooft coupling constant for $U(N)$
Chern-Simons theory on $L_{\mu}$.
In this formula, $(\widehat{t},\widehat{s})$ denote closed string flat coordinates, which are
related to the classical complexified K\"ahler parameters $(t,s)$
by the mirror map. For the present model the relation between
$(\widehat{t},\widehat{s})$ and $(t,s)$
has been discussed in \cite{AViv}.
By analogy $\widehat{t}_{op}$ denotes an open string flat coordinate corresponding
to $t_{op}$ \cite{AVi,AKV,LM,Mi,Mii}. To conclude this section,
note that the topological A model amplitudes are
independent of $\mu$ by standard decoupling arguments,
so we will drop the subscript $\mu$ from now on.
In order to check the duality predictions we need exact expressions
for both terms in (\ref{eq:dualityAi}), which will be worked out in the
next sections.
\section{Closed string amplitudes}
\label{part3}
In this section we consider closed string A model amplitudes
on $X$. The partition function ${\CF}^{cl}(t,s,g_s)$ has been
computed in \cite {AViv} up to terms depending on $t$. The strategy is
to first compute the genus zero partition function using local mirror
symmetry, and then write down a complete formula by interpreting
the answer in terms of BPS invariants \cite{GVii,kkv,MM}.
Note that this method is
specific to the present model; in general one cannot derive all the
higher genus amplitudes from the genus zero expression and BPS constraints.
The resulting expression is\footnote{From now on we will denote the flat
coordinates by $(t,s)$ dropping the symbol $\widehat{}$ .
Although this notation is potentially ambiguous,
the meaning should be clear from the context. In particular, topological
amplitudes will always be written in terms of flat coordinates.}
\begin{equation}\label{eq:closedA}
{\CF}^{cl}(t,s,g_s) = \sum_{n\geq 1} \left[ {e^{-ns}\over n(2\sin(ng_s/2)^2)}
+{e^{-n(s+t)}\over n(2\sin(ng_s/2)^2)}\right] +
\left(\hbox{$t-$dependent terms}\right).
\end{equation}
Note that throughout this paper we will mainly consider truncated
partition functions, i.e. we will drop the typical polynomial terms
which occur in the low genus expansion.
In order to run a precision test of the duality 
(\ref{eq:dualityAi}), we need to find the remaining $t$-dependent terms.
This can be done using local mirror symmetry by analogy with \cite{AViv}
The local mirror manifold of the A model described in section two is
a hypersurface $W$ in $\IC^2 \times (\IC^*)^2$ which can be written
in terms of flat coordinates as \cite{AViv}
\begin{equation}\label{eq:locmirrA}
zw=(1-e^{-u})(1-e^{u-t})-e^{-v}(1-e^{u-t-s})
\end{equation}
where $(z,w)$ are coordinates on $\IC^2$ and $(e^{-u},e^{-v})$
are single valued open string flat coordinates on $(\IC^*)^2$
\cite{AVi,AKV,GJT,LM,Mi,Mii}.
The terms in (\ref{eq:closedA}) have been found by computing the first derivative
$\partial_s F_0(t,s)$ of the genus zero partition function as a
classical period in the local mirror geometry \cite{CKYZ,HV,HIV}.
We have
\begin{equation}\label{eq:locmirrB}
\partial_sF_0(t,s)= \int_{t+s}^{\Lambda} v(u)du
\end{equation}
where
\begin{equation}\label{eq:locmirrC}
v(u) = \log\left[{1-e^{u-t-s}\over (1-e^{-u})(1-e^{u-t})}\right].
\end{equation}
The integral is taken along the semi-infinite contour $\gamma_s$
represented in fig. 2, which can be thought as the projection of a
lagrangian three-cycle onto the $(u,v)$ plane. Since this cycle has
infinite volume, we have to introduce an infrared cut-off $\Lambda$.
Evaluation of (\ref{eq:locmirrC}) results in
\begin{equation}\label{eq:genuszeroA}
\partial_s F_0(t,s)=
-\sum_{n\geq 1} \left[{e^{-n(s+t)}\over n^2}+{e^{-ns}\over n^2}\right]
\end{equation}
up to polynomial and $\Lambda$-dependent terms.
As explained in \cite{AViv} this suggests the existence of two stable
BPS states with charges
$[C_s],[C_s+C_t]\in H_2(X,\IZ)$. The terms in (\ref{eq:closedA}) represent the
contributions of these two states to the BPS invariants.

\begin{figure}[ht]
    \centering
    \scalebox{1}{\includegraphics[angle=0]{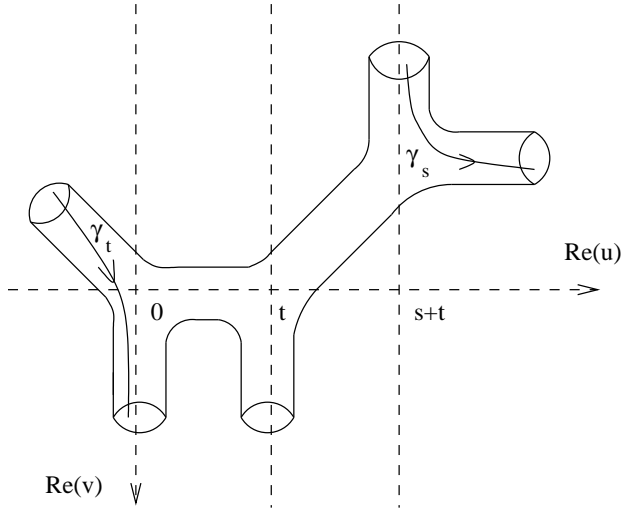}}
    \caption{Mirror Riemann surface}\label{fdhha}
    
\end{figure}

In order to detect eventual $t$-dependent terms, one has to perform
a similar computation of $\partial_tF_0(t,s)$. The relevant contour is
$\gamma_t$ represented in fig. \ref{fdhha}. We have

\begin{equation}\label{eq:locmirrD}
\partial_tF_0= \int_{-\infty}^0 v(u)du =\int_{0}^\infty v(-u)du
\end{equation}
with $v(u)$ given by (\ref{eq:locmirrC}). Since we are mainly interested in truncated
amplitudes, it suffices to compute $\partial^2_{t}F_0(s,t)$ in order
to avoid convergence issues. Then we are left with an easy calculation

\begin{equation}\label{eq:locmirrE}
\partial^2_tF_0(s,t)  = \int_{0}^\infty
\left[{e^{-u-t-s}\over 1-e^{-u-t-s}}-{e^{-u-t}\over
1-e^{-u-t}}\right]du
= \sum_{n\geq 1} \left[{e^{-n(t+s)}\over n}-
{e^{-nt}\over n}\right].
\end{equation}

Combining (\ref{eq:genuszeroA}) and (\ref{eq:locmirrE}) we obtain the following
expression for the truncated genus zero amplitude
\begin{equation}\label{genuszeroB}
F_0(t,s)= \sum_{n\geq 1} \left[{e^{-ns}\over n^3}+{e^{-n(t+s)}\over n^3}
-{e^{-nt}\over n^3}\right].
\end{equation}
Reasoning by analogy with \cite{AViv}, this form of the genus zero
amplitude predicts an extra term in the expression of the closed
string partition function
\begin{equation}\label{eq:closedB}
{\CF}^{cl}(t,s,g_s) = \sum_{n\geq 1} \left[ {e^{-ns}\over n(2\sin(ng_s/2)^2)}
+{e^{-n(s+t)}\over n(2\sin(ng_s/2)^2)}
- {e^{-nt}\over n(2\sin(ng_s/2)^2)}\right].
\end{equation}
A couple of remarks are in order at this point. The expressions (\ref{eq:closedA}),
(\ref{eq:closedB}) are conjectural at this stage since they have not been
confirmed by explicit A model closed string computations. While the terms
in (\ref{eq:closedA}), have a clear interpretation in terms of BPS invariants,
for the third term in (\ref{eq:closedB}), such an interpretation is more subtle.
The problem is that in the class $[C_t]$
there is a family of rational curves  with parameter space $\IC$,
which is noncompact. Therefore
it is not clear how to define the BPS invariants or the
Gromov-Witten invariants in this case.
A possible approach is to choose an appropriate compactification of
$X$ and then take a large volume limit. The resulting invariants
will be well defined, but they may be dependent on the particular
compactification chosen. This is a drawback, since such compactifications
are far from unique. At the same time, we should keep in mind that
the above B model computation gives an answer (at least at genus zero),
without making such a choice. So a legitimate question is whether
this computation has an A model counterpart.
We will propose below a partial answer, leaving a more conceptual
approach for future work.

\begin{figure}[h]
     \begin{center}
     \scalebox{1}{\includegraphics{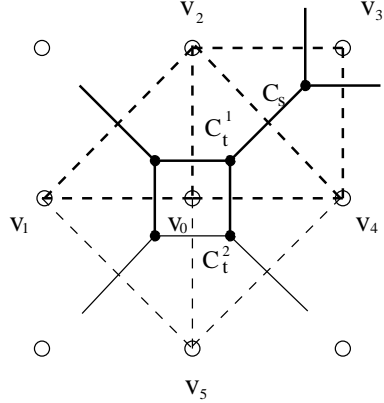}}
     \caption{A Toric Compactification of $X$.}\label{edhha}
     \end{center}
     \end{figure}
To begin with, note that we can compactify this model to a local
Calabi-Yau variety containing a Hirzebruch surface $\IF_0$
and a transverse $(-1,-1)$ curve. This is represented in fig. \ref{edhha}.
The class $[C_t]$ is the fiber class of $\IF_0$, and $C_s$ can be identified
with the transverse curve to $\IF_0$. This is a flopped version of the
local $dP_2$ model. In this new model, the family of curves in class $[C_t]$
has a compact parameter space $\IP^1$, and the BPS invariants in the class
$d[C_t]$ are very simple \cite{kkv}. We have $n_1^0=-e(\IP^1)=-2$ and $n_d^r=0$
for all other values of $(d,r)$, where $e(\IP^1)$ is the Euler character of
$\IP^1$. There are two stable BPS states whose wavefunctions
are harmonic representatives of $H^{0,0}(\IP^1)$ and respectively
$H^{1,1}(\IP^1)$. In order to recover our original model $X$, we have
to take an infinite volume limit of the base $\IP^1$ of $\IF_0$.
In this limit, the harmonic function in $H^{0,0}(\IP^1)$ becomes
non-normalizable, hence the corresponding BPS state is lifted from
the spectrum. We are left with a single BPS state corresponding to
the harmonic $(1,1)$-form on $\IP^1$, which can be kept normalizable
in this limit. Taking into account the sign as well, we obtain precisely
the contribution in (\ref{eq:closedB}).
To conclude our discussion of closed string amplitudes, note that we
can also find an interpretation of (\ref{eq:closedB}) in Gromov-Witten theory.
Namely, using the same compactified model, one can try to compute the
Gromov-Witten invariants in the class $n[C_t]$ by localization
with respect to an $(\IC^*)^2$ action on $\IF_0$ \cite{CKYZ}.
The $(\IC^*)^2$ action leaves two fibers invariant, which will be
denoted by $C_t^1$, $C_t^2$; $C_t^1$ passes through the intersection
point with $C_s$, while $C_t^2$ is a fiber at infinity from the point
of view of $X$. The fixed locus of the $(\IC^*)^2$ action on
${\mo_g}(X^c, n[C_t])$ consists accordingly of two classes 
of components corresponding to maps onto $C_t^1$ and respectively $C_t^2$.
Moreover, by symmetry, it is easy to see that the two classes of 
components have equal contributions to the integral representation
of Gromov-Witten
invariants for all $(g,n)$. Since the integral in question is taken over a
compact moduli space, the sum of the two contributions must be a rational
number, hence each individual contribution must also be a rational number.
In the decompactification limit, the fiber at infinity is effectively
removed, and we are left with a single $(\IC^*)^2$-invariant component.
Therefore we can unambiguously define a closed string expansion on $X$ by
simply taking the contribution of a single fixed component of
${\mo_g}(X^c, n[C_t])$. For genus zero, and low enough
degree, it can be checked that this contribution is $-{1\over n^2}$
as expected. We conjecture that this procedure also gives
the expected answer for all $(g,n)$, but we will not try to prove it here.

To summarize the main point of this section we have presented
compelling evidence that the complete closed partition
function on $X$ is given by (\ref{eq:closedB}). We will show in the next section,
that this expression is also in very good agreement with geometric duality
predictions.

\section{Open String Amplitudes and The Duality Map}
\label{openstring}

Let us consider now open string A model amplitudes on the deformation
space $Y$ with $N$ D-branes wrapped on $L$.
According to \cite{EWii},
the target space effective action for $N$ D-branes wrapping
$L$ is $U(N)$ Chern-Simons gauge theory. This theory has played
a central role in large $N$ geometric duality for the conifold,
starting with
\cite{GViii}. The novelty in the case under consideration is that
the Chern-Simons theory is corrected by open string instanton effects.
Such situations have been anticipated in section 4.4. of \cite{EWii},
where the instanton effects have been elegantly interpreted as
nonlocal Wilson loop corrections to the Chern-Simons action.
Suppose for simplicity that we have a single holomorphic disc $D$
in $Y$, with boundary on $L$.
Schematically, the full effective action can be written as
\begin{equation}\label{eq:effactA}
S(A) = S_{CS}(A) + F_{inst}(g_s,t_{op}, V)
\end{equation}
where $t_{op}$ is an open string flat coordinate\footnote{We have dropped again
the symbol $\widehat{}$ in the notation of flat coordinates.},
$g_s = {2\pi \over k+N}$ is the renormalized Chern-Simons
coupling constant and $V= \hbox{Pexp}\int_{\Gamma}A$ is the
holonomy of the $U(N)$ connection around the boundary $\Gamma$
of $D$, regarded as a knot in $L$.
For large K\"ahler parameters, the instanton corrections can be treated
perturbatively from the Chern-Simons point of view.
Moreover, in the present paper we are interested in a large $N$
't Hooft expansion so that the
open string free energy takes the form
\begin{equation}\label{eq:freenA}
{\CF}_{op}(t_{op},\lambda,g_s) ={\CF}^{0}_{op}(\lambda,g_s)
+\hbox{ln}\left\langle e^{F_{inst}(g_s,t_{op},V)}\right\rangle,
\end{equation}
where $\lambda = Ng_s$ is the 't Hooft coupling constant
and ${\CF}^{0}_{op}(\lambda, g_s)$ is the Chern-Simons free energy.
In order to evaluate (\ref{eq:freenA}) we need an exact expression for the
open string instanton corrections $F_{inst}(g_s,t_{op}, V)$. We
will give here a schematic treatment, leaving some formal details
for the next section. Recall that in section two we have
constructed a holomorphic disc $D$ embedded in $Y$ with boundary
on $L$. We reproduce for convenience the local expression of the
embedding map from (\ref{eq:holdiscA}) 

\begin{equation}\label{eq:holdiscB}
\begin{aligned}[b]
&U_1:&~~&\lambda(t)=t,& &v(t)={\mu\over t},& &x(t) = 0,& &y(t)=0\\
&U_2:&~~&\rho(t')=t',& &v(t')=\mu t',& &x(t')=0,& &u(t') = 0
\end{aligned}
\end{equation}
 where $D$ is the disc $\{|t|\geq
\mu^{1/2}\}=\{|t'|\leq \mu^{-1/2}\}$ in $\IP^1$ with affine
coordinates $(t,t')$. The boundary of $D$ is mapped to the sphere
$L$, which is defined by 
\begin{equation}\label{eq:sphereF}
\begin{aligned}[b]
&U_1:& &\lambda ={\overline{v}},& &x ={\overline{y}} \cr 
&U_2:& &\rho {\overline{v}} =1,& &x = {\overline{\rho}} {\overline{u}}.
\end{aligned}
\end{equation}
Note that the second coordinate patch $U_2$ covers the disc
$D$, but it does not cover the entire sphere $L$. In fact,
equation (\ref{eq:sphereF}) shows that $U_2$ covers $L$ with the circle $\{
\lambda=v=0, |x|=|y|=\mu^{1/2} \}$ removed, since $v,\rho$ are not
allowed to vanish on $L\cap U_2$. This shows that $L\cap U_2$ is
diffeomorphic to $S^1\times \IR^2$ and the boundary of $D$ is a
section of this cylinder which can be identified with the circle
$|t'|=\mu^{-1/2}$. Moreover, the normal bundle to $D$ in $Y$ can
be identified with a trivial rank two bundle on $D$ with
coordinates $(x,u)$ along the fiber. These coordinates are subject
to the boundary conditions 

\begin{equation}\label{eq:boundcond}
x = {\overline \rho}{\overline u}.
\end{equation} 
This is a familiar situation since the boundary
conditions (\ref{eq:boundcond}) are identical to the ones of
\cite{KL,LS}. In particular, using their results, it follows
that the embedding $f:D\rightarrow Y$ is rigid. For completeness, note
that the boundary conditions (\ref{eq:boundcond}) define a totally real
subbundle $N_{\IR}$ of the normal bundle $N$ restricted to the
boundary $\Gamma=\partial D$. The pair $(N, N_{\IR})$ forms a
Riemann-Hilbert bundle with generalized Maslov index $\mu(N,
N_\IR)=-1$. By the double construction, the group of global
sections $H^0(D,\Gamma;N, N_\IR)$ is zero. We will show in the
next section that $D$ is a generator of $H_2(Y,L;\IZ)=\IZ$.
By analogy with the closed string situation, the open string
instanton numbers should be defined in terms of intersection theory
on the moduli space of maps from bordered Riemann surfaces to the
pair $(Y,L)$ \cite{KL}.
The homotopy classes of open string maps $f:\Sigma_{g,h} \rightarrow Y$ with
$f(\partial \Sigma_{g,h}) \subset L$ are classified by the class
$\beta = d[D]$, with $d\in \IZ$. We denote by $M_{g,h}(Y,L;d[D])$ the
moduli space of stable open string maps to the pair $(Y,L)$ such that
$f_*[\Sigma_{g,h}]= d[D]$, as defined in \cite{KL}. This space should
have a suitable compactification $\om_{g,h}(Y,L;d[D])$. At the moment,
very little is known about the structure of the compactified space
$\om_{g,h}(Y,L;d[D])$. However, in the particular case considered here,
this space has at least certain disconnected components which are
familiar from the work of \cite{KL, LS}.
The components in question are isomorphic to moduli spaces of
multicovers of the disc $D$, which are disconnected components
of $\om_{g,h}(Y,L;d[D])$ because $D$ is rigid. In addition to the degree
$d$, the multicovers of $D$ are also characterized by the winding numbers
$n_\alpha$, $\alpha=1,\ldots, h$ of the $h$ boundary components.
Therefore we have disconnected components ${\om}_{g,h}(D,\Gamma;d,n_\alpha)$.
According to \cite{KL} these should be thought of as orbifold spaces with boundary.
Note that at this point we do not know if ${\om}_{g,h}(Y,L;d[D])$ has
other disconnected components in addition to ${\om}_{g,h}(D,\Gamma;d,n_\alpha)$.
The open string Gromov-Witten invariants $N_d$ for maps of degree $d$
should be defined in terms of a virtual fundamental class of degree zero
on $\om_{g,h}(Y,L;d[D])$ \cite{KL} whose construction is not known at
the present stage. Assuming that such a construction exists, the invariants
$N_d$ will receive contributions $N_{d, n_\alpha}$
from all disconnected components
$\om_{g,h}(D,\Gamma;d,n_\alpha)$ which have been evaluated in
\cite{KL,LS}. In addition, we may have contributions from other
disconnected components of $\om_{g,h}(Y,L;d[D])$, which are beyond our
control at the present stage. We will however present an argument in
a later section suggesting that such extra contributions are absent.
For the time being, we assume that this is the case, and write down
the answer found in \cite{KL,LS}. Since the spaces
$\om_{g,h}(D,\Gamma;d,n_\alpha)$ have boundary, the invariants
$N_{d,n_\alpha}$ depend on the choice of boundary conditions, which
in the present case amounts to the choice of a framing of $\Gamma$
\cite{KL,LS,OV}, i.e. the choice of a homotopy class of sections
of $N_\IR$. In the canonical framing, we have the following instantonq
expansion
\begin{equation}\label{eq:instcorrB}
F_{inst}(g_s,t_{op},V) = i\sum_{d=1}^{\infty}{e^{-dt_{op}}\over 2d\hbox{sin}{dg_s\over 2}}\tr V^d.
\end{equation}
One can write similar, although more complicated expressions for an
arbitrary framing, but we will not give more details here. However, it
would be very
interesting to understand how duality works at arbitrary framing.
We thank Mina Aganagic and Cumrun Vafa for clarifying discussions
on this point.
The target space effective action (\ref{eq:effactA}) becomes
\begin{equation}\label{eq:effactB}
S(A) = S_{CS}(A) + i\sum_{d=1}^{\infty}
{e^{-dt_{op}}\over 2d\hbox{sin}{dg_s\over 2}}\tr V^d.
\end{equation}
Performing a analytic continuation along the lines of 
 \cite{MRD,GVi,GViii,VP}, the 't Hooft expansion of the
Chern-Simons free energy can be set in the form
\begin{equation}\label{eq:freenB}
{\CF}^{0}_{op}(g_s,\lambda)=\sum_{n=1}^\infty {e^{in\lambda}\over
n\left(2\hbox{sin}{ng_s\over 2}\right)^2}
\end{equation}
where $\lambda = Ng_s$ is the 't Hooft coupling constant. 
Recall that throughout this paper we consider truncated string amplitudes,
so we have dropped a polynomial and logarithmic piece. Those terms
are well understood in the context of geometric duality \cite{GViii}.

In order to finish the computation we have to evaluate the expectation
value of $e^{F_{inst}(g_s,t_{op},V)}$. For an unknot with the canonical
framing we have \cite{OV}
\begin{equation}\label{eq:unknotA}
\left\langle \tr V^{k_1}\ldots \tr V^{k_l}\right\rangle =
\left\langle \tr V^{k_1}\right\rangle \ldots
\left\langle \tr V^{k_l}\right\rangle
\end{equation}
for any positive integers $k_1, \ldots, k_l$.
Using this property and expanding the exponential, we find
\begin{equation}\label{eq:unknotB}
\hbox{ln}\left\langle e^{F_{inst}(g_s,t_{op},V)}\right\rangle=
i\sum_{d=1}^\infty
{e^{-dt_{op}}\over 2d\hbox{sin}{dg_s\over 2}}\left\langle
\tr V^d\right\rangle.
\end{equation}
The expectation values $\left\langle \tr V^d \right\rangle$
have been evaluated in \cite{OV} for arbitrary $d$, with the result
\begin{equation}\label{eq:unknotC}
\left\langle \tr V^d \right\rangle =
i{{e^{-id\lambda/2} - e^{id\lambda/2}}\over 2\hbox{sin}{dg_s\over 2}}.
\end{equation}
Collecting all the results, we obtain the following expression for
the open string free energy
\begin{equation}\label{eq:freenBi}
{\CF}_{op}(t_{op},\lambda,g_s) = \sum_{n=1}^\infty {e^{in\lambda}\over
n\left(2\hbox{sin}{ng_s\over 2}\right)^2}+
\sum_{d=1}^\infty
{e^{-d(t_{op}-{i\lambda\over 2})}-e^{-d(t_{op}+{i\lambda\over 2})}
\over d\left(2\hbox{sin}{dg_s\over 2}\right)^2}.
\end{equation}
This expression is to be compared with the closed string free energy worked
out in section three.

\subsection{Comparison with Closed String and Duality Map}
Recall that in the previous section we found
the closed string free energy on $X$ to be
\begin{equation}\label{eq:closedstr}
{\CF}_{cl}(g_s, s, t) = \sum_{n=1}^\infty
{1\over n \left(2\hbox{sin}{ng_s\over 2}\right)^2}
\left(e^{-ns}+e^{-n(t+s)}-e^{-nt}\right).
\end{equation}
This formula is remarkably similar to (\ref{eq:freenBi}), except for the
different dependence of the instanton factors on K\"ahler moduli.
However, it is straightforward to see that we will obtain a precise
agreement if we conjecture a duality map relating
the closed and open string K\"ahler parameters by
\begin{equation}\label{eq:dualityA}
s=-i\lambda,\qquad t= t_{op} +{i\lambda\over 2}.
\end{equation}
Note that the relation $s=-i\lambda$ differs by a sign from the 
duality map obtained in \cite{GVii}. Changing the sign in the duality map
corresponds to a flop in the closed string geometry. One can check
that the alternative duality map $s=i\lambda$ 
corresponds to a different large radius limit of the model dicussed in
section two, namely $\hbox{Re}(t)>0$, $\hbox{Re}(s)<0$.

At this point, one may wonder what is the physical interpretation 
of the half integral shift of the second K\"ahler parameter. 
In the large volume regime,
$t,t_{op}$ differ from the classical K\"ahler parameters by exponentially
small corrections. Hence, as discussed in more detail at the end of
section five, a classical geometric reasoning would
predict a relation of the form $t=t_{op}$, which is obviously
in contradiction with (\ref{eq:dualityA}).

On the other hand, the shift (\ref{eq:dualityA}) appears to be
a perturbative quantum correction in Chern-Simons theory. Such a
correction would not be visible at tree level in the field theory,
which is consistent with the treatment of \cite{AViv}. From a string theory
point of view, we can think of this shift as a nonperturbative
correction to the open string flat coordinates induced by degenerate
open string instantons.
Recall that the Chern-Simons perturbation expansion has been
interpreted as a sum over virtual instantons at infinity
in \cite{EWii} (section 4.2.) These are essentially open string Riemann
surfaces which degenerate to trivalent graphs geodesically embedded in
$L$. In the present context, we can have a partial degeneration
of a surface $\Sigma_{g,h}$ to another surface $\Sigma_{g',h'}$
mapped to the disc $D$ and a trivalent graph
with external legs ending on the boundary of $D$.
For example, the degeneration of a sphere with three holes is represented in Fig.4.
\begin{figure}[h]
     \begin{center}
     \scalebox{1}{\includegraphics{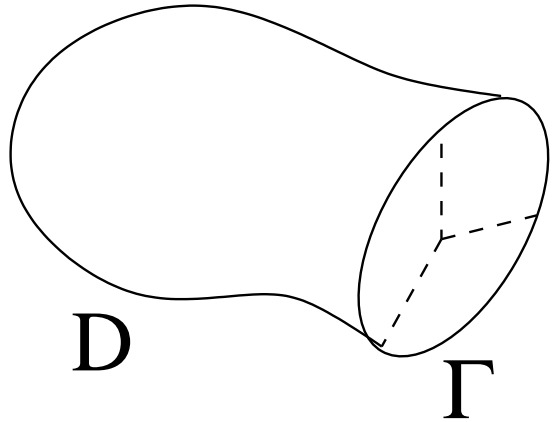}}
     \caption{A degenerate open string instanton of type $(g,h)=(0,3)$}\label{dedhha}
     \end{center}
     \end{figure}
In the field theory limit, the  sum over degenerate instantons is equivalent
to a sum over trivalent graphs (in double line notation) with external
legs ending on $\Gamma$.
This is nothing else but the perturbative expansion of the Wilson line
vacuum expectation values in Chern Simons theory performed above.
\section{The Geometry of Open String Maps}
\label{gosm}

We have shown so far that the open string instanton computations
are in very good agreement with the predictions of large $N$ duality
provided that one sums only over multicovers of a rigid disc $D$
in $Y$. A legitimate question at this point is if there are
additional contributions from other components of the
moduli space of open string maps which may spoil this agreement.
This is the problem we would like to address in this section,
by developing a more formal approach to open string morphisms
to the pair $(Y,L)$.

The main difficulty in answering this question is the lack of explicit
constructions for generic open string maps in this background.
This results in a very poor understanding of the moduli space
$M_{g,h}(Y,L;d[D])$ and its compactification.
We can considerably simplify our task by adopting the strategy of
\cite{GZ,KL}, which means we can restrict our considerations to
open string maps invariant under a certain torus action on $Y$
which preserves $L$. In principle, this should allow us to settle the
issue of extra
contributions without a detailed knowledge of the moduli space.

In order to gain a better control on the geometry, we first compactify the
hypersurfaces $Y_\mu$ by taking a projective
closure of the ambient variety $Z$.
Recall that $Z$ is isomorphic to the total space of $\CO(-1)\oplus 2\CO$ over
$\IP^1$, which can be represented as a toric variety
\begin{equation}\label{eq:toricDf}
\begin{array}{llllll}
& Z_1 &Z_2 & X & Y & V \\
\IC^* & 1 & 1 & 0 & -1 & 0
\end{array}
\end{equation}
with disallowed locus $\{Z_1=Z_2=0\}$.
The family $\CY/\Delta$ can be described as a family of hypersurfaces in
$Z$ determined by
\begin{equation}\label{famhyper}
Z_1V+XYZ_2^2 = \mu Z_2.
\end{equation}
The local coordinates $(x,y,v,\lambda)$ and
$(x,u,v,\rho)$ are standard affine toric coordinates for (\ref{eq:toricDf}).
For example, $\lambda={Z_1/Z_2},\ x={X},\ y={YZ_2},\ v={V}$.
The (relative) projective closure of $Z$ is the compact toric variety
$\bz = \left(\IC^6\setminus F\right)/(\IC^*)^2$
determined by
\begin{equation}\label{eq:toricD}
\begin{array}{ccccccc}
 &Z_1 &Z_2 & X & Y & V & W\cr
\IC^* & 1 & 1 & 0 & -1 & 0 & 0 \cr
\IC^* & 0 & 0 & 1 & 1 & 1 & 1\cr
\end{array}
\end{equation}
with disallowed locus $F=\{Z_1=Z_2=0\}\cup \{X=Y=V=W=0\}$.
Note that $\bz\simeq \IP\left(3\CO\oplus \CO(-1)\right)$ over $\IP^1$.
Then the projective completion of the family (\ref{famhyper}) is a family
$\bcy/\Delta$ of compact hypersurfaces in $\bz$ given by
\begin{equation}\label{eq:defC}
Z_1VW +XYZ_2^2 = \mu Z_2W^2.
\end{equation}
Let $\by$ denote a generic fiber of this family (as noted before, we drop
the subscript $\mu$ with the understanding that $\mu$ is fixed at some
real positive value.) We denote by
$Y{\buildrel i\over \inj} \by {\buildrel j\over \inj} \bz$
the obvious embedding maps.  The divisor at infinity on $\by$,
$\zeta_\infty$ is defined as
the pull back of the Cartier divisor $W=0$ on $\bz$,
so that $Y=\by\setminus \zeta_\infty$.
There is a subtlety related to this compactification, namely the
variety $\by$ is singular along the locus
\begin{equation}\label{eq:sinlocusA}
\{Z_2=V=W=0\} \cup \{Z_1=X=Y=W=0\}
\end{equation}
which is entirely contained in the divisor at infinity. In principle, one
should blow-up these singularities in order to have a good control
over the geometry. However, for the present application this step is
not really necessary since we can reduce our problem to questions about
curves on $\bz$, which is smooth. After completing the argument, it will
become clear that blowing up $\by$ along the singular locus at infinity
does not affect the conclusion.

For future reference, note that in terms of homogeneous coordinates
the local coordinates on $Y$ are given by
\begin{equation}\label{eq:loccoordB}
\begin{aligned}[b]
& U_1:& &\qquad \lambda={Z_1\over Z_2},& &\qquad x={X\over W},& &
\qquad y={YZ_2\over W},& &\qquad v={V\over W}\\
& U_2:& &\qquad \rho={Z_2\over Z_1},& &\qquad x={X\over W},& &
\qquad u={YZ_1\over W},& & \qquad v={V\over W}.\\
\end{aligned}
\end{equation}
In the following, we will also need local coordinates at infinity
defined in the open set $U_3=\{Z_2\neq 0, V\neq 0\}$
\begin{equation}\label{eq:inftycoord}
\lambda = {Z_1\over Z_2},\qquad x'={X\over V},\qquad
y'={YZ_2\over V},\qquad v'={W\over V}.
\end{equation}

Let us record some details on the geometry of $\bz$.
Since $\bz$ is toric and given by (\ref{eq:toricD}), its Picard group has
rank two and it is generated by toric divisors. We can pick a
system of generators $(\xi_1, \xi_2)$ determined by the divisor
classes
\begin{equation}\label{eq:torgenA}
\xi_1:\ (Z_1=0)=(Z_2=0), \qquad
 \xi_2:\ (X=0)=(V=0)=(W=0).
\end{equation}
Note that $Y=0$ defines a divisor in the class
$\xi_2-\xi_1$ and, using an explicit set of generators of the fan,
we obtain the relations
\begin{equation}\label{eq:intersA}
\begin{aligned}[b]
&\xi_1^2 =0,& & \xi_2^3(\xi_2-\xi_1)=0\\
&\xi_1\xi_2^3 =1,& & \xi_2^4=1.
\end{aligned}
\end{equation}
If we regard $\bz$ as a $\IP^3$ fibration over
$\IP^1$, $\xi_1$ is the class of a fiber, and $\xi_2$ is the class
of a relative hyperplane section. The Mori cone of $\bz$ is
generated by curve classes corresponding to 3-cones in the toric
fan determined by the data (\ref{eq:toricD}). From (\ref{eq:intersA}) it is clear
that a basis of the Mori cone is given by
\begin{equation}\label{eq:moriA}
\eta_1=\xi_2^2(\xi_2-\xi_1),\qquad \eta_2=\xi_1\xi_2^2.
\end{equation}
Let us choose some convenient representatives of these
curve classes
\begin{equation}\label{eq:moriB}
\eta_1:\ X=Y=V=0,\qquad \eta_2:\ Z_1=X=Y=0.
\end{equation}

After these preliminary remarks, we are ready to discuss open
string maps. As mentioned in the first paragraph of this section,
we restrict our considerations to open string morphisms which are
fixed points of a certain torus action. In the present context, we
will consider a torus action on $Y$ induced by an action on $\by$
which preserves the divisor at infinity. There is a natural
$(S^1)^4$ action on $\bz$ with weights 
\begin{equation}\label{eq:toractA} 
\begin{array}{lccccccccccc}
&Z_1& &Z_2& &X& &Y& &V& &W\\ 
&\lambda_1& &0& &\lambda_2& &\lambda_3& &\lambda_4& &0.
\end{array}
\end{equation}
 The subgroup $(S^1)^2\subset
(S^1)^4$ defined by $\lambda_1+\lambda_4=0$ and
$\lambda_2+\lambda_3=0$ preserves the hypersurface $\by$ and the
sphere $L$. It is also clear that the divisor $\zeta_\infty$ is
invariant, therefore we obtain a well defined $(S^1)^2$-action on
$Y$. For localization purposes, it suffices to consider a diagonal
subgroup $T\subset (S^1)^2$ acting on  $(Y,L)$. In the coordinate
patch $U_1$, this action reads 
\begin{equation}\label{eq:toractB} 
\lambda \rightarrow
e^{-i\lambda_1\theta} \lambda,\qquad x\rightarrow e^{-i\lambda_2\theta}x
\qquad y\rightarrow e^{i\lambda_2\theta}y\qquad v\rightarrow
e^{i\lambda_1\theta}v.
\end{equation}

Since the action of $T$ preserves $L$, it induces an action on the
moduli space of open string morphisms with lagrangian boundary
conditions on $L$. Our strategy is to find the fixed points of
this action  subject to a homology constraint. First note that any
$T$-invariant stable map $f:\Sigma_{g,h}\rightarrow Y$ with $f(\partial
\Sigma_{0,1})\subset L$ must have the following form
\cite{GZ,KL,LS}. The domain $\Sigma_{g,h}$ is of the form
$\Sigma^0_{g}\cup \Delta_1\cup \Delta_2 \cup\ldots \Delta_h$ where
$\Sigma_{g}^0$ is a prestable curve of arithmetic genus $g$ with
$h$ marked points $p_1, \ldots p_h$ and $\Delta_1,\ldots,
\Delta_h$ are discs attached to $\Sigma^0_{g}$ by identifying the
origin of each disc with a point $p_\alpha$, $\alpha=1,\ldots, h$.
The map ${f}_{|\Sigma^0_g}:\Sigma^0_g\rightarrow Y$ must be a
$T$-invariant stable map to $Y$, and ${f}_{|\Delta_\alpha}:
\Delta_\alpha \rightarrow Y$ must be a $T$-invariant map to a disc in $Y$
(with boundary) on $L$.

There is however a significant difference between our model and
those of \cite{GZ,KL,LS}. In those cases, although the target
space $Y$ is a noncompact Calabi-Yau threefold, the maps actually
take values in a compact submanifold thereof, which is simply a
disc in \cite{KL,LS} and a projective plane in \cite{GZ}. In
our case we have certain components of the moduli space which
consist of multicovers of a rigid disc, but we do not know a
priori that these are all the components. Therefore we cannot a
priori assume that the map $f:\Sigma_{g,h} \rightarrow Y$ takes values in
a compact submanifold of $Y$. The correct treatment of this
situation is to work in a relative setting, namely we should
consider open string maps to the pair $(\by, L)$ with prescribed
order of contact along the divisor at infinity $\zeta_\infty$. For
this, one should blow-up the singularities of $\by$ at infinity,
and consider maps to the resulting smooth three-fold. This would
be an open string version of relative Gromov-Witten theory which
will not be pursued here in detail. In principle, in this physical
situation, one should consider open string morphisms with order of
contact zero at infinity, and there would be very subtle questions
related to the compactification of the moduli space. It will
eventually become clear that for the present purposes we do not
need to develop a full theory along these lines; it suffices to
extend our search to $T$-invariant open string maps to the pair
$(\by, L)$ subject to the homology constraint
$f_*[\Sigma_{g,h}]=d[D]$. Since $\by$ has singularities at
infinity, conceptually, we can think of maps $f:\Sigma_{g,h} \rightarrow
\by$ as maps $f:\Sigma_{g,h}\rightarrow \bz$ with boundary conditions on
$L$, such that\footnote{Recall that $j:\by\rightarrow \bz$ denotes the
embedding. There is a subtlety related to this point, since we may
have different homology classes in $H_2(\by,L;\IZ)$ which are
mapped to the same homology class in $H_2(\bz, L,\IZ)$. This can
be taken into account by a careful treatment.}
$f_*[\Sigma_{g,h}]=j_* d [D]$ and the image of $f$ lies in $\by$.
This is the point of view we will take below, when we refer to
open string maps to $\by$. In the end, we will show that the
structure of the fixed locus is such that there are no other
contributions to the open string invariants besides those
considered in the previous section.

Given the special structure of $T$-invariant open string maps,
the problem reduces to
finding $T$-invariant maps $f:\Sigma_{0,1}\rightarrow \by$
with lagrangian boundary conditions along $L$.
We claim that any
such map can be extended to a $T$-invariant map
${\overline f}: {\overline \Sigma}_0 \rightarrow \by$ from a smooth
genus zero curve ${\overline \Sigma}_0\simeq \IP^1$ to $\by$.
This further reduces the problem to searching for
$T$-invariant rational curves on $\by$ which intersect $L$.

In order to justify this claim, recall that $L$ is defined in
the coordinate patch $(x,y,v,\lambda)$ by
\begin{equation}\label{eq:sphereB}
v\lambda +xy=\mu,\qquad \lambda = {\overline v}, \qquad
x={\overline y}.
\end{equation}
Take a tubular neighborhood $Q_\epsilon$ of $L$ in $Y$ (hence
also in $\by$) of the form $Q_\epsilon = Y \cap B^8_\epsilon$,
where $B^8_\epsilon$ is the eight-ball
\begin{equation}\label{eq:ballA}
|\lambda|^2 + |v|^2 + |x|^2 + |y|^2 \leq 2(\mu^2+\epsilon^2)^{1/2}
\end{equation}
where $\epsilon\in \IR_+$ (recall that $\mu$ is also taken
real and positive throughout this paper.)
According to \cite{HC}, $Q_\epsilon$ is a complex manifold with boundary
diffeomorphic to $B^3\times S^3$. The boundary
$S=\partial Q_\epsilon$
is diffeomorphic to $S^2 \times S^3$. Moreover, since $L$ is lagrangian,
it follows that $S$ is a contact hypersurface with respect to the
symplectic form induced from $Y$ \cite{ EGH, JE}.
If we take $\epsilon$ small enough, for any map $f:\Sigma_{0,1}\rightarrow \by$
with $f(\partial \Sigma_{0,1})\subset L$, $f(\Sigma_{0,1})\cap Q_\epsilon$
is a small cylinder $\Xi$
embedded in  $Q_\epsilon$ with the two boundary components
mapped to $L$ and respectively $S$.
Given the local form of the $T$-action (\ref{eq:toractB}), if $f:\Sigma_{0,1}\rightarrow \by$ is $T$-invariant, 
there are only four such cylinders that can occur this way. We have
\begin{equation}\label{eq:cyl}
\begin{aligned}[b]
&\Xi_1:& &\mu^{1/2}\leq |\lambda|\leq\left[{(\mu^2+\epsilon^2)^{1/2}+\epsilon}\right]^{1/2},& & v={\mu\over \lambda},& &x=y=0\\
&\Xi_2: & &\left[{(\mu^2+\epsilon^2)^{1/2}-\epsilon}\right]^{1/2}\leq |\lambda|\leq \mu^{1/2},& & v={\mu\over \lambda},& &x=y=0\\
&\Xi_3: & &\mu^{1/2}\leq |x|\leq\left[{(\mu^2+\epsilon^2)^{1/2}+\epsilon}\right]^{1/2},&  &y={\mu\over x},& &\lambda=v=0\\
&\Xi_4: & &\left[{(\mu^2+\epsilon^2)^{1/2}-\epsilon}\right]^{1/2}\leq |x|\leq \mu^{1/2},& &y={\mu\over x},& &\lambda=v=0.\\
\end{aligned}
\end{equation}

In each of these cases, we can find a suitable extension of $f$ to a map
$\of : {\overline \Sigma}_0 \rightarrow \by$.
For example let us consider $\Xi_1$.
Since $f$ is $T$-invariant, we can find a local coordinate
$t$ on $f^{-1}(\Xi_1)$ and a
positive integer $k$ such that $f$ is locally given by
\begin{equation}\label{eq:locmap}
\lambda(t) = t^k,\qquad v(t)={\mu\over t^k}, \qquad x(t)=y(t)=0.
\end{equation}
Note that in this parameterization, $f^{-1}(\Xi_1)$ is isomorphic
to the annulus
\begin{equation}\label{eq:annulus}
\mu^{1/2k}\leq |t| \leq
\left[{(\mu^2+\epsilon^2)^{1/2}+\epsilon}\right]^{1/2k}.
\end{equation}
Now we can define a map from the disc
\begin{equation}\label{eq:disc}
D(\epsilon, k):\qquad 0\leq |t|\leq
\left[{(\mu^2+\epsilon^2)^{1/2}+\epsilon}\right]^{1/2k}
\end{equation}
to $\by$ which in the local coordinates at infinity (\ref{eq:inftycoord})
reads
\begin{equation}\label{eq:inftymap}
\lambda = t^k,\qquad v'(t) = {1\over \mu} t^k, \qquad x(t)=y(t)=0.
\end{equation}
Here we are forced to work in the coordinate patch at infinity
since $v\rightarrow \infty$ as $t\rightarrow 0$ in (\ref{eq:locmap}). In particular, the origin
of the disc $D(\epsilon, k)$ is mapped to the point at infinity
$P=[0,1,0,0,1,0]$.
Then we can glue the disc $D(\epsilon, k)$ to $\Sigma_{0,1}$
along the annulus (\ref{eq:annulus}) obtaining a smooth rational curve
${\overline \Sigma}_0$
and the map $f:\Sigma_{0,1}\rightarrow \by$ extends by (\ref{eq:inftymap})
to a map $\of : {\overline \Sigma}_0\rightarrow \by$.
By $T$-invariance, the image of this map has to be a rational curve
in $\by$ preserved by $T$ and passing through the point at infinity
$P$. The only curve on $\by$ satisfying these conditions is
\begin{equation}\label{eq:fixedcurveA}
C_1:\qquad Z_1V = \mu WZ_2, \qquad X=Y=0.
\end{equation}
By writing (\ref{eq:fixedcurveA}) in local coordinates, it follows that $C_1$
intersects
$L$ along a circle which divides it into two discs
with boundary on $L$. One of them $D_1$ is the disc $D$
considered in the previous section, while the
second one $D_1^\prime$ is a disc in $\by$ with origin at $P$.
The invariant map $f:\Sigma_{0,1}\rightarrow \by$ is a $k:1$ cover of $D$.

Similar considerations apply to the other three cases in (\ref{eq:cyl}).
For the second case, we obtain again the curve (\ref{eq:fixedcurveA}), but the
roles of $D_1$ and $D_1^\prime$ are reversed. We now obtain a $k:1$ cover
of the disc $D_1^\prime$. For the remaining cases, we find a $T$-invariant
curve
\begin{equation}\label{eq:fixedcurveB}
C_2:\qquad XYZ_2 =\mu W^2, \qquad Z_1=V=0
\end{equation}
which is divided by $L$ into two discs $D_2, D_2^\prime$ in $\by$.
By contrast with the previous situation, both
$D_2$ and $D_2^\prime$ intersect the divisor at infinity at $Q=[0,1,0,1,0,0]$
and respectively $R=[0,1,1,0,0,0]$. Moreover, the invariant map
$f:\Sigma_{0,1}\rightarrow \by$ is a $k:1$ cover of $D_2$ and respectively
$D_2^\prime$.
To summarize this discussion, we conclude that an invariant map from a
disc to $\by$ with boundary conditions on $L$ must be a multicover
of one of the four discs $D_1, D_1^\prime, D_2, D_2^\prime$ found above.
Note that except $D_1=D$, the other three discs have points at infinity,
hence they are not contained in the noncompact hypersurface $Y$.
For this reason, one might be tempted at this point to rule out all
fixed points
consisting of maps with components along $D_1^\prime, D_2$ or $D_2^\prime$,
keeping only maps with components along $D$. However, since we do not
really understand the structure of the moduli space, we should proceed
with more caution here.

We will show next that the discs
$D_1^\prime, D_2$ or $D_2^\prime$ are in fact ruled out by the
homology constraint $f_*[\Sigma_{g,h}]=d[D]$. The idea is to push
forward homology classes to $\bz$, which is smooth, and use the
structure of the Mori cone.
First, note that using a standard exact sequence argument, we have
an isomorphism
\begin{equation}\label{eq:homisom}
0\rightarrow H_2(\bz, \IZ){\buildrel \alpha \over \rightarrow}  H_2(\bz, L;\IZ)\rightarrow 0.
\end{equation}
Therefore, it suffices to compute the homology classes
$\alpha^{-1}[D_1],\ldots, \alpha^{-1}[D_2']$ in terms of
the generators $\eta_1, \eta_2$ of the Mori cone defined in (\ref{eq:moriB}).
This is not quite straightforward, since the discs are rigid, and
one cannot measure their homology class
by using intersection theory as in the case of holomorphic curves.
Instead we have to use the following deformation argument.
Suppose we deform the sphere $L$ in $\bz$ be changing the value of
$\mu$ to $\mu'<\mu$. In this paragraph we restore the subindex $\mu$ for
$Y, L$ in order to keep track of the $\mu$-dependence. Then the disc
$D_{1\mu}$ also changes to $D_{1\mu'}$ which can be obtained from
$D_{1\mu}$ by gluing in a small cylinder, which is fillable in $\bz$.
This shows that $\alpha^{-1} [D_{1\mu}]=\alpha^{-1}[D_{1\mu'}]$ for
any $\mu, \mu'$, and the same is true for the other three discs.

The advantage of this approach is that we can deform
to $\mu=0$, such that the sphere $L_\mu$ shrinks to zero size. In this limit,
the discs become holomorphic curves on $\bz$
whose homology classes in $H_2(\bz, \IZ)$ can be easily determined from
the algebraic equations. Let us consider for example
the discs $D_1, D_1^\prime$. In the limit
$\mu=0$, the defining equations (\ref{eq:fixedcurveA}) of $C_1$ specialize
to
\begin{equation}\label{eq:limitA}
Z_1V=0, \qquad X=Y=0.
\end{equation}
Therefore $C_1$ specializes to a reducible curve with components
$X=Y=V=0$ and respectively $Z_1=X=Y=0$, which are precisely the generators
(\ref{eq:moriB}) of $H_2(\bz, \IZ)$. The two discs $D_1, D_1^\prime$ are deformed
in this limit to these two components of $C_1$, therefore we find
\begin{equation}\label{eq:homclsA}
[D_1] = \eta_1, \qquad [D_1^\prime]=\eta_2.
\end{equation}
By a similar reasoning
we also find $[D_2]=[D_2^\prime]=\eta_2$.
Note that this deformation argument shows that the symplectic area of 
the disc $D$ is the same as that of the curve $C_t$ after transition. 
Therefore at classical level, $t=t_{op}$ as noted at the end of section two. 

Now we can determine the general structure
of a $T$-invariant open string morphisms subject to the constraint
$f_*[\Sigma_{g,h}]=d[D]$, with $d$ a positive integer. Since $\eta_1$,
$\eta_2$ are generators of the Mori cone, it follows that
for any such fixed point $f|_{{\Delta_\alpha}}:\Delta_{\alpha} \rightarrow \by$
must be a multicover of $D$. The other discs are indeed ruled out
by homology constraints since one cannot have effective curves $C$
on $\bz$ so that $d\eta_1=[C] + d'\eta_2$ for $d, d'>0$. Moreover,
by the same argument, the closed
curve $\Sigma_g^0$ is mapped either to a point or to a $T$-invariant
rational curve in the class $\eta_1$. Now, one can check that the only
$T$-invariant curves on $\by$ in this class are the sections defined
by $X=Y=W=0$, $X=V=W=0$ and $V=Y=W=0$. These are all included in the divisor
at infinity $\zeta_\infty$, and they are disconnected from the
invariant disc $D$. Since the image of any map has to be connected,
it follows that we cannot have fixed open string maps with components
along the above curves. This leaves only maps that contract the
curve $\Sigma_g^0$ to a point, while mapping $\Delta_\alpha$
to $D$. These are precisely the fixed points in the multicover moduli
space of $D$, considered in \cite{KL,LS}, whose contributions
have been taken into account in the previous section. After this rather
lengthy analysis, we can conclude that this is the complete answer.
 
\bigskip\noindent
{\bf Acknowledgements:} 

We are very grateful to Mina Aganagic and Cumrun Vafa for
illuminating discussions and suggestions and to Ron Donagi and
Tony Pantev for collaboration on a related project.  We would also
like to thank Bobby Acharya, Michael Douglas, John Etnyre,
Albrecht Klemm, Andrew Kresch, Marcos Mari\~no, John McGreevy,
Harald Skarke  and Lisa Traynor for very stimulating
conversations. We owe special thanks (and lots of tiramis\`{u})
to Corina Florea for invaluable 
help with the LaTeX conversion of the original draft. 
The work of D.-E. D. has been supported by DOE
grant DOE-DE-FG02-96ER40959; A.G. is supported in part by the  NSF Grant
DMS-0074980.

\bibliographystyle{alpha}

\begin{thebibliography}{AAAA}
 
\bibitem{AAHV}{B.
Acharya, M. Aganagic, K. Hori and C. Vafa, ``Orientifolds, Mirror
Symmetry and Superpotentials'', hep-th/0202208.}
\bibitem{AVi}{M. Aganagic and C. Vafa, ``Mirror Symmetry,
D-Branes and Counting Holomorphic Discs'', hep-th/0012041.}
\bibitem{AKV}{M. Aganagic, A. Klemm and C. Vafa, ``Disk Instantons,
Mirror Symmetry and the Duality Web'', Z. Naturforsch. {\bf A 57}
(2002) 1, hep-th/0105045.}
\bibitem{AVv}{M. Aganagic and
C. Vafa, ``Mirror Symmetry and a $G_2$ Flop'', hep-th/0105225.}
\bibitem{AViv}{M. Aganagic and C. Vafa, ``$G_2$ Manifolds, Mirror
Symmetry and Geometric Engineering'', hep-th/0110171.}
\bibitem{B}{J. Blum, ``Calculation of Nonperturbative
Terms in Open String Models'', hep-th/0112039.}
\bibitem{CKYZ}{T.-M. Chiang, A. Klemm,
S.-T. Yau and E. Zaslow, ``Local Mirror Symmetry: Calculations and
Interpretations'', ATMP {\bf 3} (1999) 495, hep-th/9903053.}
\bibitem{HC}{H. Clemens, ``Double Solids'', Adv.
Math. {\bf 47} (1983) 107.}
\bibitem{MRD}{M.R. Douglas, ``Chern-Simons-Witten Theory as a
Topological Fermi Liquid'', hep-th/9403119.}
\bibitem{EGH}{Y. Eliashberg, A.
Givental and H. Hofer, ``Introduction to Symplectic Field
Theory'', math.SG/0010059.}
\bibitem{JE}{J.B.
Etnyre, ``Symplectic Convexity in Low Dimensional Topology'',
Topology Appl. {\bf 88} (1998) 3.}
\bibitem{GVi}{R. Gopakumar and C. Vafa, ``Topological Gravity as Large
$N$ Topological Gauge Theory'', ATMP {\bf 2} (1998) 413,
hep-th/9802016.}
\bibitem{GVii}{R. Gopakumar and C. Vafa, `` M-Theory and Topological Strings--I'',
hep-th/9809187;  `` M-Theory and Topological Strings--II'', hep-th/9812127.}
\bibitem{GViii}{R. Gopakumar and C. Vafa, ``On the Gauge Theory/Geometry
Correspondence'', ATMP {\bf 3} (1999) 1415, hep-th/9811131.}
\bibitem{GJT}{S. Govindarajan, T. Jayaraman and T. Sarkar, ``Disc
Instantons in Linear Sigma Models'', hep-th/0108234.}
\bibitem{GZ}{T. Graber and E. Zaslow, ``Open-String Gromov-Witten
Invariants: Calculations and a Mirror Theorem' '', hep-th/0109075.}
\bibitem{MG}{M. Gross, ``Topological Mirror Symmetry'',
math.AG/9909015.}
\bibitem{HV}{K. Hori and C. Vafa, ``Mirror Symmetry'', hep-th/0002222.}
\bibitem{HIV}{K. Hori, A. Iqbal and C. Vafa, ``D-Branes and Mirror Symmetry'',
hep-th/0005247.}
\bibitem{IK}{A.
Iqbal and A.-K. Kashani-Poor, ``Discrete Symmetries of the
Superpotential and Calculation of Disk Invariants'',
hep-th/0109214.}
\bibitem{KKLMi}{S.
Kachru, S. Katz, A. Lawrence and J. McGreevy, ``Open String
Instantons and Superpotentials'', Phys. Rev. {\bf D62} (2000)
026001, hep-th/9912151.}
\bibitem{KKLMii}{S. Kachru, S. Katz, A.
Lawrence and J. McGreevy, ``Mirror Symmetry for Open Strings'',
hep-th/0006047.}
\bibitem{KL}{S. Katz and C.-C. M. Liu, ``Enumerative Geometry of Stable
Maps with Lagrangian Boundary Conditions and Multiple Covers of
the Disc'', ATMP {\bf 5} (2001) 1, math.AG/0103074.}
\bibitem{kkv}{A. Klemm, S. Katz and C. Vafa, ``M-Theory,
Topological Strings and Spinning Black Holes'', ATMP {\bf 3}
(1999) 1445, hep-th/9910181.}
\bibitem{LMi}{J.M.F. Labastida and M. Mari\~no, ``Polynomial Invariants for
Torus Knots and Topological Strings'',  Commun. Math. Phys. {\bf 217} (2001)
423, hep-th/0004196.}
\bibitem{LMV}{J.M.F. Labastida, M. Mari\~no and C. Vafa, ``Knots,
Links and Branes at Large $N$'', JHEP {\bf 11} (2000) 007, hep-th/0010102.}
\bibitem{LMii}{J.M.F. Labastida and M. Mari\~no, ``A New Point of View in the
Theory of Knot and Link Invariants'', math.QA/0104180.}
\bibitem{LM}{W. Lerche and P. Mayr, ``On ${\cal N}=1$ Mirror Symmetry
for Open Type II Strings'', Hep-th/0111113.}
\bibitem{LV}{N.-C. Leung and C. Vafa, ``Branes and
Toric Geometry'', ATMP {\bf 2} (1998) 91, hep-th/9711013.}
\bibitem{LS}{J. Li and Y.S. Song, ``Open String Instantons and Relative Stable
Morphisms'', hep-th/0103100.}
\bibitem{MM}{M. Mari\~no and G. Moore, ``Counting higher genus 
curves in a Calabi-Yau manifold'',Nucl.Phys. {\bf B543} (1999) 592,  hep-th/9808131.}
\bibitem{MV}{M. Mari\~no and C. Vafa, ``Framed Knots at Large $N$'',
hep-th/0108064.}
\bibitem{Mi}{P. Mayr, ``${\cal N}=1$ Mirror Symmetry and Open/Closed String
Duality'', hep-th/0108229.}
\bibitem{Mii}{P. Mayr, ``Summing Up Open String Instantons and ${\cal
N}=1$ String Amplitudes'', hep-th/0203237.}
\bibitem{OV}{H. Ooguri and C. Vafa, ``Knot Invariants and Topological
Strings'', Nucl. Phys. {\bf B 577} (2000) 419, hep-th/9912123.}
\bibitem{VP}{V. Periwal, ``Topological Closed String Interpretation of
Chern-Simons Theory'', Phys. Rev. Lett. {\bf 71} (1993) 125,
hep-th/9305115.}
\bibitem{RS}{P. Ramadevi and T. Sarkar, 
``On Link Invariants and Topological String Amplitudes'', 
Nucl. Phys. {\bf B600} (2001) 487, hep-th/0009188.}
\bibitem{CVi}{C. Vafa, ``Extending Mirror Conjecture
to Calabi-Yau with Bundles'', hep-th/9804131.}
\bibitem{sts}{C. Vafa, ``Superstrings and Topological
Strings at Large $N$'', J. Math. Phys. {\bf 42} (2001) 2798,
hep-th/0008142.}
\bibitem{EWii}{E.
Witten, ``Chern-Simons Gauge Theory as a String Theory'',
``The Floer Memorial Volume'', H. Hofer, C.H. Taubes, A. Weinstein
and E. Zehnder, eds, Birkh\"auser 1995, 637,
hep-th/9207094.}


\end{thebibliography}

\end{document}